\documentclass[aps,prl,twocolumn,reprint,superscriptaddress,english]{revtex4-1}
\usepackage[utf8]{inputenc}
\usepackage[T1]{fontenc}
\usepackage{bm}
\usepackage{graphicx}
\usepackage{float}
\usepackage{wrapfig}
\usepackage[dvipsnames]{xcolor}
\usepackage{placeins}
\usepackage[linktocpage=true,colorlinks=true,pdfborder={0 0 0},linkcolor=blue,
citecolor=blue,filecolor=yellow,urlcolor=blue,bookmarks,pdfauthor={},]{hyperref}
\usepackage{amsmath}
\usepackage{gensymb}
\usepackage{nicefrac}
\usepackage[utf8]{inputenc}

\newcommand{\LasVegas}{Department of Physics \&{} Astronomy, University of Nevada Las Vegas, Las Vegas, Nevada 89154, USA}
\newcommand{\LasVegasChem}{Department of Chemistry \&{} Biochemistry, University of Nevada Las Vegas, Las Vegas, Nevada 89154, USA}
\newcommand{\HPCAT}{HPCAT, X-ray Science Division, Argonne National Laboratory, Illinois 60439, USA}
\newcommand{\RochesterME}{Department of Mechanical Engineering, University of Rochester, Rochester, New York 14627, USA}
\newcommand{\RochesterPhys}{Department of Physics \&{} Astronomy, University of Rochester, Rochester, New York 14627, USA}
\newcommand{\ETH}{Department of Earth Sciences, ETH Zürich, Zürich 8025, Switzerland}
\newcommand{\Empa}{Centre for X-ray Analytics, Empa – Swiss Federal Laboratories for Materials Science and Technology, \"{U}berlandstraße 129, 8600 D\"{u}bendorf, Switzerland.}

\begin{document}

\author{G.~Alexander~Smith}
 \affiliation{\LasVegasChem}
 \author{Ines E. Collings}
 \affiliation{\Empa}
 \author{Elliot~Snider}
 \affiliation{\RochesterME}
 \author{Dean~Smith}
 \affiliation{\HPCAT}
 \author{Sylvain Petitgirard}
 \affiliation{\ETH}
\author{Jesse Smith}
 \affiliation{\HPCAT}
 \author{Melanie White}
 \affiliation{\LasVegas}
  \author{Elyse Jones}
 \affiliation{\RochesterME}
  \author{Paul Ellison}
 \affiliation{\LasVegas}
\author{Keith~V.~Lawler}
 \affiliation{\LasVegasChem}
\author{Ranga~P.~Dias}
 \affiliation{\RochesterME}
 \affiliation{\RochesterPhys}
\author{Ashkan~Salamat}
  \email{ashkan.salamat@unlv.edu}
 \affiliation{\LasVegas}

\date{\today{}}

%\title{Phases and metastable states of superconducting photo-induced carbonaceous sulfur hydride below 1\,Mbar}\AS{ranga doesnt like this title}

\title{Lower pressure phases and metastable states of superconducting photo-induced carbonaceous sulfur hydride}
\begin{abstract}

Room-temperature superconductivity was recently discovered in carbonaceous sulfur hydride (C-S-H) close to 3\,Mbar. 
%revealed density-driven quantum behavior in a hydrogen-rich system.
We report significant differences in the superconducting response of C-S-H, with a maximum $T_{C}$ of 191(1)\,K,  below a 1\,Mbar. 
Variations in intensity of the C-H Raman modes reveal carbon content can vary between crystals synthesized with the same photo-induced method.
Synchrotron single crystal x-ray diffraction identifies polymorphism with increasing degrees of covalency.
These unique metastable states are highly sensitive to thermodynamic pathways.

\end{abstract}

\maketitle

In the search for superconductivity at ambient conditions, the current highest reported transition temperature, $T_{C}$, is for a carbonaceous sulfur hydride (C-S-H) material with a $T_{C}$ of 288\,K at 267\,GPa.\cite{CSH}
C-S-H belongs to a class of high-hydrogen content materials realized at over a megabar of pressure, whose investigation was ignited by the discovery of a 203\,K $T_{C}$ at 155\,GPa in hydrogen sulfide.\cite{Drozdov2015,Duan2014,Errea2016}
These hydrogen-rich materials are of interest due to Ashcroft's prediction that a hydrogen dominant alloy will act as a precompressed version of elemental hydrogen with lower transition pressures into the metallic and superconducting states.\cite{PhysRevLett.92.187002}
The precompressed hydrogen alloys are expected to share the same strong electron-phonon coupling, high Debye temperature, high hydrogen-related density of states at the Fermi level, and thus high-$T_{C}$ superconductivity as predicted for dense metallic hydrogen.
The Wigner-Huntington phase of dense atomic metallic hydrogen has long been predicted to be a phonon-mediated type-II high-$T_{C}$ superconductor,\cite{WH1935,Ashcroft68} with dense molecular metallic hydrogen superconducting at even higher temperatures due to enhancements coming from electron-electron interactions.\cite{PhysRevLett.78.118}

\begin{figure}
    \centering
    \includegraphics{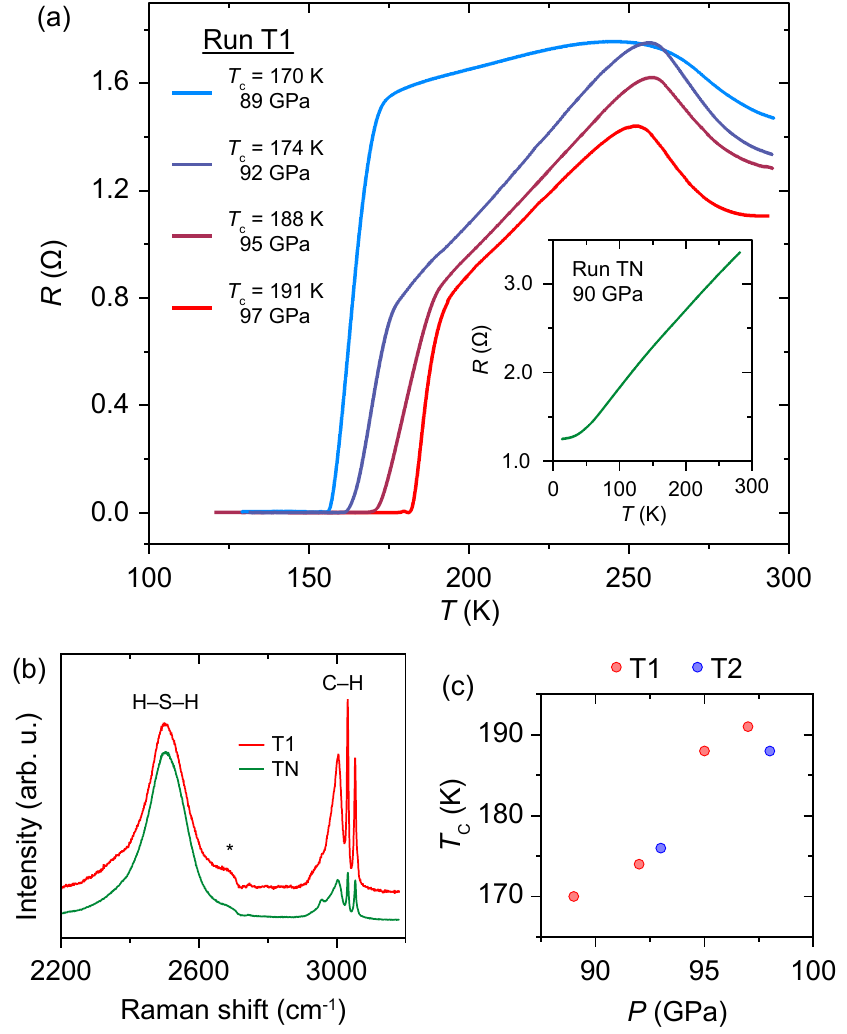}
    \caption{
    (a) The resistance response of a C-S-H crystal (Run T1) between pressures of 89 and 97\,GPa are shown as a function of temperature.
    Results show the resistance drops to zero at 191\,K at 97\,GPa implying the transition to a superconducting state.
    (Inset) The metallic resistance response of Run TN at 90\,GPa.
    (b) Raman spectra of runs T1 and TN at 4.0 GPa, showing variation in C-H : H-S-H intensities between syntheses
        The feature marked with an asterisk (*) is the second-order Raman scattering from diamond.
    (c) Evolution of critical temperature, $T_{C}$, with pressure for runs T1 and T2 %(Data for T2 can be found in Fig. \ref{SI:trans} in the Supplemental Materials) in this study.
    }
  \label{fig:trans}
\end{figure}

C-S-H, like hydrogen sulfide, is a covalent superhydride, wherein hydrogen is believed to form extended bonding networks with the other elements present.\cite{doi:10.1146/annurev-conmatphys-031218-013413,PhysRevLett.126.117003,Belli2021}
This is opposed to the other main class of metal superhydides, which present as cage-like clathrate structures.
The original theoretical attempts to identify candidate structures for C-S-H suggested a stoichiometry of CSH$_7$. 
These structures consist of a molecular methane (CH$_4$) unit encapsulated in an H$_3$S perovskite-like sublattice with the rotational orientation of the CH$_4$ unit relative to the H$_3$S sublattice determining the symmetry.\cite{PhysRevB.101.134504,PhysRevB.101.174102}
However, the most favorable structures were found to have insufficiently large $T_{C}$ values to justify their assignment as C-S-H. 
Other studies probed the possibility of carbon substitution for a sulfur in a hydrogen sulfide lattice,\cite{GE2020100330,hu2020carbondoped}
which obtained better agreement with experimental $T_{C}$ values.
However, there are concerns over the validity of the structures owing to the inability of virtual crystal approximation (VCA) to capture the structural deformations arising from the doping.\cite{PhysRevB.104.064510,wang2021little}
Additional stoichiometries have been suggested, with the leading candidates being methane substituted hydrogen sulfide lattices with a C:S ratio closer to those of the VCA simulations than the 50\% of the CSH$_7$ structures.\cite{wang2021little}

C-S-H was first synthesized via photochemistry from elemental precursors at 4 GPa.
Raman spectroscopy revealed a rich phase diagram at low pressures (10s of GPa).\cite{CSH}
This draws analogues to the H$_{2}$S--H$_{2}$ and CH$_{4}$--H$_{2}$ binary systems, which form extended van der Waals structures comprising the constituent molecules.\cite{Somayazulu1996,PhysRevLett.107.255503}
In the absence of thermal annealing, as in the initial experiments, the compression pathway likely leads to a metastable state.
Metastable states are known to be unreliably predicted using crystal structure prediction (CSP) methodologies. 
C-S-H has recently been experimentally synthesised using an alternative method, reacting elemental sulfur and  methane-hydrogen fluid mixtures.\cite{goncharov2021synthesis}
In principal, this method permits greater control of methane concentration, although reported C-H Raman modes are comparably weak, and whether such routes lead to high $T_C$ superconducting states has yet to be studied.   

To better understand the lower pressure phases of C-S-H, and motivated by the recent discovery of anomalous superconducting properties in YH$_6$,\cite{https://doi.org/10.1002/adma.202006832} we investigate C-S-H in the sub-megabar regime to further understand the synthesis chemistry and the consequences of the thermodynamic pathways to metallization and superconductivity.
We present electrical transport measurements in this previously unexplored pressure regime that reveal a remarkably high $T_{C}$ in some crystals, raising the question as to how these macroscopic quantum states emerge over such dramatically different pressure-temperature ranges.
Using synchrotron single crystal X-ray diffraction (SC-XRD), the phase progression of C-S-H was determined to be $I4/mcm$ $\rightarrow$\ $C2/c$
$\rightarrow$\ $I4/mcm$, below a megabar, demonstrating an evolution from 
van der Waals to hydrogen bonding dominating ordering. 
We propose that the carbon content in C-S-H produced by photo-chemistry methods varies in each crystal synthesised.
That variation directly affects the quantum properties for superconductivity, with subtle differences in packing densities.

Crystals of C-S-H are synthesized using the procedure of \citet{CSH} (full details in the Supplemental Materials).\footnote{See Supplemental Material at http://link.aps.org/supplemental/DOI for additional experimental and computational details.} 
Ball-milled mixtures of elemental carbon and sulfur with dimensions about 15\% of the diamond culet are placed into the sample chamber with a ruby sphere.\cite{ruby} The samples used in these runs are significantly larger, by roughly 3-10 times, using culets varying from 250 down to 100\,$\mu{}$m than in \cite{CSH}.
Gas phase H$_2$ is loaded at 3\,kbar.\cite{doi:10.1063/1.5048316}
Samples are then pressurized to 3.7--4.0\,GPa and excited for several hours using a 514\,nm laser with power ranging from 10 and 150\,mW depending on sample response.
Raman spectroscopy confirms the transformation into C-S-H with the presence of C-H, S-H, and H-H Raman modes at $\sim$4\,GPa.\cite{CSH}
Larger crystals are less homogeneous than crystals formed using 30\,$\mu{}$m culets needed for ultra high pressures, as we observe variations of the appearance of C-H stretching modes to be spatially dependant using Raman spectroscopy.
Samples are pressurized to 10\,GPa after transformation and confirmation by Raman spectroscopy to avoid decomposition.
Electrical transport measurements are carried out according to \citet{CSH}. 
Synchrotron single crystal X-ray diffraction (SC-XRD) measurements were conducted at Sector 16, ID-B, HPCAT at the Advanced Photon Source ($\lambda$ = 0.34453\,$\text{\AA}$). 

%Unlike previous multi-megabar measurements performed on C-S-H by \citet{CSH}, 

%the samples used in these runs are significantly larger, by roughly 3-10 times, using culets varying from 250 down to 100\,$\mu{}$m.
%This likely results in crystals that are less homogeneous than crystals formed using 30\,$\mu{}$m culets needed for ultra high pressures, as we observe variations of the appearance of C-H stretching modes to be spatially dependant using Raman spectroscopy.

Figure~\ref{fig:trans} shows the results of the electrical transport measurements performed on three pristine C-S-H crystals at sub-megabar regimes.
138\,GPa was the lowest pressure examined previously owing to a notion that the phase diagram and properties of C-S-H would be similar to hydrogen sulfide.
In two separate runs, we observe a maximum $T_{C}$ of 191(1)\,K at 97(5)\,GPa (T1, plots in Figure~\ref{fig:trans}a and red points in Figure~\ref{fig:trans}c), and 188(1)\,K at 98(5)\,GPa (T2, blue points in Figure~\ref{fig:trans}c), at roughly half the previously reported pressures for similar transition temperatures in \cite{CSH} (Fig \ref{fig:SI-DTc} in the SI). 
The onset of superconductivity begins at 170(1)\,K and 89(5)\,GPa (trend shown in Figure~\ref{fig:trans}c).
The shape of the superconducting dome shown in Figure~\ref{fig:trans}c being distinct from the previous data suggests this dome comes from a different phase of C-S-H than the maximal $T_C$=288\,K.
Also observed in Run T1 is the previously noted behavior of C-S-H to exhibit increasingly narrow $\Delta{}T$/$T_C$ as a function of increasing pressure and $T_C$, displaying the minimum $\Delta{}T$/$T_C$ of 0.0373 at 97\ GPa. (data can be found in Figure \ref{fig:SI-DTc} and Table \ref{tab:SI2} in the Supplemental Materials).
This superconducting transition is compared to the response of another C-S-H crystal, at a comparable pressure of 90\,GPa, which, though metallized as evidenced by its  decreasing resistance on cooling, does not becomes superconducting down to 10(1) K (Fig.~\ref{fig:trans} inset, run TN). 
The corresponding Raman spectrum of the superconducting C-S-H crystal exhibits stronger signal strength of the C-H stretching frequencies relative to the H-S-H mode (1.16 C-H/S-H).
This ratio is comparably stronger than the same Raman features of Run TN when normalized, which exhibited a C-H/S-H ratio of only 0.27.
The Raman spectrum shown in Fig \ref{fig:trans}b for run TN is more similar those reported previously.\cite{CSH, goncharov2021synthesis}
%All of the three transport measurements ended due to diamond failure at the highest pressures reported.

All of the $R\left(T\right)$ features for the different pressures measured for the superconducting C-S-H crystal from Run T1 feature a turning point around 250\,K (Fig. ~\ref{fig:trans}a).
A similar shape for $R\left(T\right)$ was observed recently in dense hydrogen ($\geq$\,320\,GPa) when it was cooled from phase V to the (semi-)metallic phase III.\cite{eremets2021metallization}
Above 250\,K, C-S-H exhibits the temperature response of a finite gap system, whereas just below 250\,K the temperature response is metallic.
This behavior at 250\,K likely results from either a structural or electronic phase transition.
An electronic transition would likely not be be accompanied by a change in symmetry, and a structural transition in a hydride material might also be indistinguishable using XRD if the heavy atom sublattice does not re-order, as is the case for the $R3m$ to $Im\overline{3}m$ transition in H$_3$S.\cite{Errea2016}
 Resistance continues to decrease with lowering temperature before a sharp drop to zero resistance as the critical temperature is crossed.
Such a difference in $T_{C}$ to that of \citet{CSH} could be expected, as their thermodynamic approach to a superconducting state began from cooling the recently reported $Im\overline{3}m$ phase observed above 160\,GPa\cite{goncharov2021synthesis} rather than phase IV.

\begin{figure}[t]
    \centering
    \includegraphics{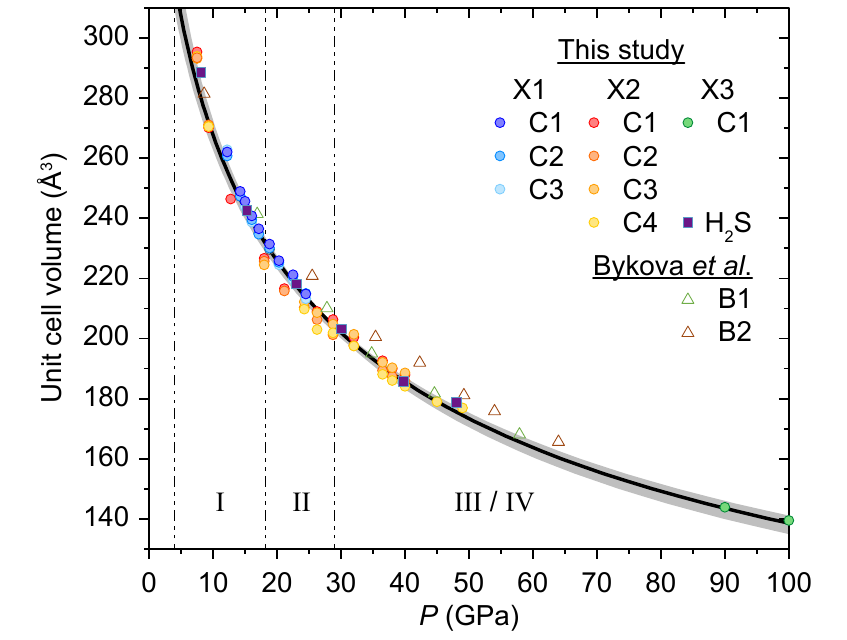}
    \caption{
    The pressure-volume relations of 8 different C-S-H crystals across three different DACs that were studied in this work are shown in circles.
    Also included are previously reported values from \citet{bykova_csh} in triangles, as well as pure hydrogen sulfide in squares.
    A 2nd Order Birch-Murnaghan equation of state was fit across all data resulting in $V_0$ = 400.573\,\AA{}$^3$ and $K_0$ = 11.028\,GPa fit parameters. 
    The gray band about the fit are for run X1 and run X2 for the high and low bands respectively.  
    Phase division included are of the I ($I4/mcm$) $\to$ II (C2/c) $\to$ III/IV ($I4/mcm$) from the Raman study in \cite{CSH}.
    Expected phases III and IV from Raman were not distinguished in XRD.
  }
    \label{fig:PV}
\end{figure}

Figure~\ref{fig:PV} shows the volume-pressure response of 8 C-S-H crystals from three separate runs.
Conical diamond with 80$^\circ$ apertures were used to enable a higher degree of completeness in SC-XRD.
We observe subtle systematic differences in volume-pressure relations across the different crystals measured at the same thermodynamic conditions.
The largest percent difference of volume was observed to be 2.9\% at 28.9(5)\,GPa in Run X2 between crystals 1 and 4.
The overall general volume trends for all of the C-S-H crystals measured are equal or lower than that of our own measurements on pure H$_2$S+H$_2$, which in turn is noticeably lower than that reported for C-S-H prepared from from mixtures of molecular H$_2$S+CH$_4$+H$_2$ gases.\cite{bykova_csh}

\begin{figure*}
    \centering
    \includegraphics[width=\textwidth]{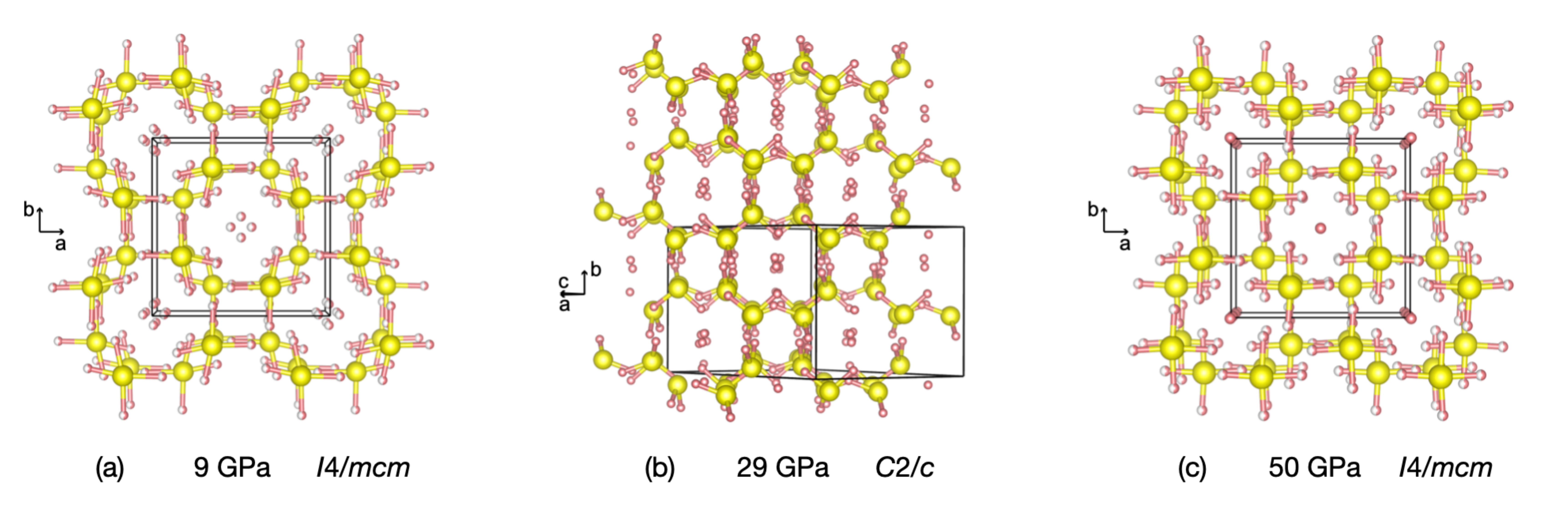}
    \caption{Structural progression of C-S-H determined by SC-XRD.}
    \label{fig:struct}
\end{figure*}

At low pressures (18\,GPa), SC-XRD measurements confirm phase I~\cite{CSH} as the Al$_{2}$Cu-type structure ($I4/mcm$) previously identified in molecular mixtures of CH$_{4}$ and H$_{2}$,\cite{Somayazulu1996} and in H$_{2}$S and H$_{2}$.\cite{PhysRevLett.107.255503,bykova_csh}
The $I4/mcm$ phase is inferred between 4 to 9\,GPa as no observed change occurred in Raman spectroscopy.
Due to insufficient C concentration or unique crystallographic placements, SC-XRD measurements are unable to resolve between C and S on the $8h$ Wyckoff positions, thus Figure~\ref{fig:struct}a displays only H$_{2}$S units on the $8h$ sites.
Applying the Bernal-Fowler "ice rules"\cite{doi:10.1063/1.1749327} to determine the H positions of these H$_{2}$S units results in partially occupied $16k$ Wyckoff positions in $I4/mcm$, and this constrains the H$_{2}$S molecular units such that they are planar within \{002\} as in \citet{PhysRevLett.107.255503}.

The 3.64\,\AA{} in-plane and 3.30\,\AA{} inter-plane nearest neighbor S--S distances are both within the hydrogen-bonded (H-bonded) dimerization distance (4.112\,\AA) of gas phase H$_2$S molecules, implying hydrogen bonding contributes to the cohesion of the lattice.\cite{https://doi.org/10.1002/anie.201808162}
A CSP study on the H--S system identified a $P1$ modification which mostly varies from the $I4/mcm$ H positions owing to the varied rotation of the molecular sub-units,\cite{Duan2014} which likely better reflect an instantaneous orientation in a thermalized sample as weak packing forces would enable the molecular sub-units to behave as weakly constrained rotors within their respective molecular volume.
Comparing several planar arrangements of the hydrogen atoms as shown in Figure~\ref{fig:struct} versus the arrangement of the $P1$ structure with density functional theory shows a $\sim$0.44\,eV enthalpic preference for a non-planar hydrogen arrangement.
This indicates C-S-H will have some non-planar arrangements of H$_2$S molecular units to create weak H-bonding between the shorter 3.30\,\AA{} interplane nearest neighbor sulfur atoms.

Molecular dynamics (MD) equilibrations which kept the lattice and sulfur positions frozen at the values determined by XRD showed H$_2$S units which were initially planar rotating such that one H atom pointed towards a S in another plane.
Although the dynamics were performed on only a single unit cell, the H$_2$S units are weakly constrained rotors going between hydrogen bonding configurations somewhat independently.
This reinforces the favorability of interplane hydrogen bonding as well as implying rotational disorder to the phase at $\geq$\,300\,K. 
The H$_2$ molecules in this structure rotated and translated completely freely within the confines of their "Cu" position in the Al$_2$Cu-type lattice, emphasizing their role as "guest" species within a guest-host structure.

At 18\,GPa, C-S-H transforms into the $C2/c$ phase shown in Figure~\ref{fig:struct}b.
This transition was observed for Runs X1 and the H$_{2}$S loading, while there is variability in the appearance of this phase for the different crystals of run X2.
A recent structural study also observed variation in the $C2/c$ phase, and its occurrence was correlated to crystals with low CH$_4$ concentrations. \cite{goncharov2021synthesis}
The $C2/c$ phase resembles a monoclinically distorted version of the $I4/mcm$ phase where the [101] direction of the $C2/c$ structure roughly corresponds to the [001] direction of the $I4/mcm$ structure.
In both cases, that direction resembles a 2-dimensional pore formed by sulfur atoms interconnected by interplane hydrogen bonding that encapsulates the H$_2$ molecules, and the views shown in Figure~\ref{fig:struct} are all oriented to look along this pore-like structure.
The positioning of the hydrogens determined by XRD are reminiscent to what is observed in the low pressure dynamics snapshots. 

Following the $C2/c$ phase, C-S-H is observed to transform back into an $I4/mcm$ structure at 29\,GPa (shown in Figure~\ref{fig:struct}c) which persists until our highest measurements at 100\,GPa. 
Our measured phase transitions by SC-XRD agree well with those reported by Raman studies.\cite{CSH} 
However, XRD does not distinguish the phase III and IV previously reported around 45\,GPa. 
This could be a reordering of hydrogen in the structure, that would be difficult to resolve in high-pressure XRD experiments.
The noted electronic transition into a metallic phase evidenced by loss or Raman signal above 65 GPa does not display evidence of a structural transition from the $I4/mcm$ solution up to our highest measurement at 100\,GPa.

The hydrogen positions of the H$_2$S units in Figure~\ref{fig:struct}c are best modeled in a planar configuration.
However, orientations with inter-plane hydrogen bonding were an order of magnitude more enthalpically favorable at 50\,GPa than 9\,GPa.
The rotations of the H$_2$S units observed with MD were highly concerted within the 50\,GPa $I4/mcm$ structure, and H$_2$S units which were initially planar remained planar in MD simulations.
These simulations show an increased importance of hydrogen bonding within the C-S-H lattice at higher pressures, indicating that molecular ordering is occurring between the low and high pressure $I4/mcm$ structures and corroborating that further ordering is the likely difference between phases III and IV.
It could be inferred from the constrained rotation that the monoclinic distortion seen in phase II likely arises from a reordering of the molecular orientations as the system densifies.

2$^{nd}$-order Birch-Murnaghan equation of state (K$_0$' fixed to 4) are fit to each crystal and phase, and the resultant values for each may be found in the supplemental materials. 
K$_0$ was found to range between 7.321 and 14.496\,GPa for Runs X1 C3 and X2 C3.
All data on phase III/IV was collected during Run X2. 
This, along with differences in the electronic response between crystals measured in this work (Figure~\ref{fig:trans}), and especially between those and crystals reported previously in \citet{CSH} produced using a similar method, suggests a remarkable variability in C-S-H generated by photochemistry under pressure (Fig~SX).
Indeed, the effects of carbon doping on the superconducting properties of H$_{3}$S have been investigated.\cite{wang2021little}

A major challenge, that demands significant improvement, for the study of C-S-H is to ensure the control of the ratio of C:S but more even challenging is the product yield and controllable concentration of the constituent elements during the photo-induced reaction.
The high-pressure photo-induced synthesis can also lead to nucleation of crystals that grow with preferred orientations resulting in difficulties manipulating crystals in a larger sample chamber.
Consequently, ensuring that not only the chemically correct crystals grow, but the ideal part of the larger crystals make full contact with the transport leads is extremely challenging.

\pagebreak 

In conclusion, new transport measurements show a transition to a superconducting state with  maximum $T_C$ of 191\,K at pressures significantly lower than previously observed.
SC-XRD marks a phase evolution of $I4/mcm$, $C2/c$, to finally $I4/mcm$ at 18 and 29\,GPa respectively.
The absence of a measurable transition from phase III to IV from earlier Raman studies indicates that the transition is likely a reordering of the hydrogens, possibly molecular species in the pores, leaving the sulfur sublattice unchanged. 
Synthesis beginning from gas phase molecular precursors presents a promising path to greater control over the chemical homogeneity of C-S-H samples.\cite{bykova_csh,goncharov2021synthesis} However, based on the path dependence metastability reported here, it needs to be confirmed that this route can produce superconducting C-S-H samples.
Ambient conditions superconductivity will eventually be achieved with hydrogen dominant alloys produced with precise chemical control.

\begin{acknowledgements}
This work supported by the U.S. Department of Energy, Office of Basic Energy Sciences under Award Number DE-SC0020303. AS and RD acknowledge Unearthly Materials for supporting this research. 
Portions of this work were performed at HPCAT (Sector 16), Advanced Photon Source (APS), Argonne National Laboratory. HPCAT operations are supported by DOE-NNSA’s Office of Experimental Sciences.
The Advanced Photon Source is a U.S. Department of Energy (DOE) Office of Science User Facility operated for the DOE Office of Science by Argonne National Laboratory under Contract No. DE-AC02-06CH11357.
\end{acknowledgements}

\end{document}